# Electric field-controlled reversible order-disorder switching of a metal tip surface

*Ludvig de Knoop[1#]\*, Mikael Juhani Kuisma[1,2#], Joakim Löfgren[1], Kristof Lodewijks[1], Mattias Thuvander[1], Paul Erhart[1]\*, Alexandre Dmitriev[1,3,4]\*, Eva Olsson[1]\*.*

[1] Department of Physics, Chalmers University of Technology, Gothenburg, Sweden

[2] Department of Chemistry, University of Jyväskylä, Jyväskylä, Finland

[3] Department of Physics, University of Gothenburg, 412 96 Gothenburg, Sweden

[4] Geballe Laboratory for Advanced Materials, Stanford University, Stanford, California 94305-4045 USA

[#] These authors contributed equally to this work.

\* Correspondence to L.D.K. (ludvig.deknoop@chalmers.se), P.E. (erhart@chalmers.se), A.D. (alexd@physics.gu.se) and E.O. (eva.olsson@chalmers.se).

KEYWORDS: Order-disorder switching, *in situ* TEM biasing, *ab initio* molecular dynamics simulation, field evaporation

While it is well established that elevated temperatures can induce surface roughening of metal surfaces, the effect of a high electric field on the atomic structure at ambient temperature has not been investigated in detail. Here we show with atomic resolution using *in situ* transmission electron microscopy how intense electric fields induce reversible switching between perfect crystalline and disordered phases of gold surfaces at room temperature. *Ab initio* molecular dynamics simulations reveal that the mechanism behind the structural change can be attributed to a vanishing energy cost in forming surface defects in high electric fields. Our results demonstrate how surface processes can be directly controlled at the atomic scale by an externally applied electric field, which promotes an effective decoupling of the topmost surface layers from the underlying bulk. This opens up opportunities for development of active nanodevices in e.g. nanophotonics and field-effect transistor technology as well as fundamental research in materials characterization and of yet unexplored dynamically-controlled low-dimensional phases of matter.

The interactions of metal surfaces with gases, liquids, solids and electromagnetic fields are of paramount importance in many fields and their applications, including, but not limited to, catalysis [1,2], plasmonic sensing [3], nano-optics [4], protein mechanics [5], biomolecular interactions [6] and nanoelectronics [7,8]. These interactions are strongly influenced by the surface structure. Such surface interactions are greatly affected by changes in the state of the surface, one example being surface roughening, which is related to surface melting [9-11], a phenomenon that has been discussed for more than a century [12,13]. Surface melting and surface roughening describes the loss of crystallinity (a disordering) of the surface layers while the underlying structure is kept crystalline [9,11]. The first experimental evidence for temperature-induced surface melting was reported in 1985 [14] and was followed by a multitude of observations at elevated temperatures [9-11,15].

Apart from elevated temperatures, the presence of an electric field of sufficient strength can change the state of the surface, e.g. in the form of field-assisted ionization and evaporation of atoms, a phenomenon used in characterization techniques such as field ion microscopy [16] and atom probe tomography [17,18]. Field evaporation has previously been studied using *in situ* transmission electron microscopy (TEM), a technique that allows simultaneous excitation and observation on the atomic scale, from an ionic liquid [19] and from carbon nanotubes for both reshaping purposes and to improve the properties for electron cold-field emission [20,21]. Cold-field emission is another effect induced by high electric fields (around 2 V/nm [22] whereas field evaporation of gold (Au) commences at around 30 V/nm), utilized in *e.g.* electron sources [23] and medical applications. Yet, little is known about the structural dynamics at ambient temperature of metallic surfaces at the atomic level at electric fields *between* the thresholds for cold-field emission and field evaporation. In this manuscript we report on the dynamics of surface Au atoms in this intermediate electric field interval. We have directly imaged the effect of the intense electric field on the atomic structure at room temperature using *in situ* TEM. We discovered that the outmost atomic layers switched from order to disorder at fields that were just below the field-evaporation field. The findings are supported by extensive *ab initio* calculations. Further on, we reverted the disordered phase back to the original crystalline form by decreasing the applied electric field, properties that are important for possible utilization in e.g. field-effect transistor and active nanodevice technologies. Subsequently, by increasing the electric field even further, we also performed and imaged field evaporation at atomic resolution.

The electric field was controlled by a biasing TEM sample holder with a piezo-driven nanomanipulator [24] [left-hand part of Fig. 1(a)] and applied to a Au nanocone [25] with a 1.7 nm tip radius at a chosen distance of 100 nm to the negative cathode. The electric field strengths were derived from finite element method (FEM) modeling taking into account atomic geometry and the applied voltage (Discussion 4 in the Supplemental Material, Fig. S3). By increasing the electrical field strengths, we observed a change in the atomic structure of the nanocone surface [cf. Figs. 1(b) and 1(c)]. Fast Fourier transformations (FFTs) of the atomic-resolution images of the apex area revealed a pattern characteristic of a crystalline face-centered cubic structure for low electrical field strengths of up to 15 V/nm [Fig. 1(d), area used from an average of 100 frames, for the actual areas of the FFT see Figs. S2(a) and S2(b)]. After increasing the electric field strength to around 25 V/nm, the outmost atomic surface layers switched from a crystalline to a non-crystalline phase and a ring-shaped feature occurred in the FFT, indicating the presence of a disordered phase [Figs. 1(e), S2(g) and S2(h)]. The ring was found to overlap with the face-centered cubic {111} diffraction spots, similarly to observations of melted aluminum [26,27], another face-centered cubic crystal. The reciprocal radius of the ring corresponds to a spatial distance of 2.4 ± 0.5 Å which matches well the distance characteristic of melted bulk Au of 2.38

Å [28]. A closer inspection suggests, however, that the disordered layers are not entirely liquidized because some underlying crystalline and ordered structure remains [Fig. 1(c)] [9,11]. In other words, the disordered state is a state of surface roughening.

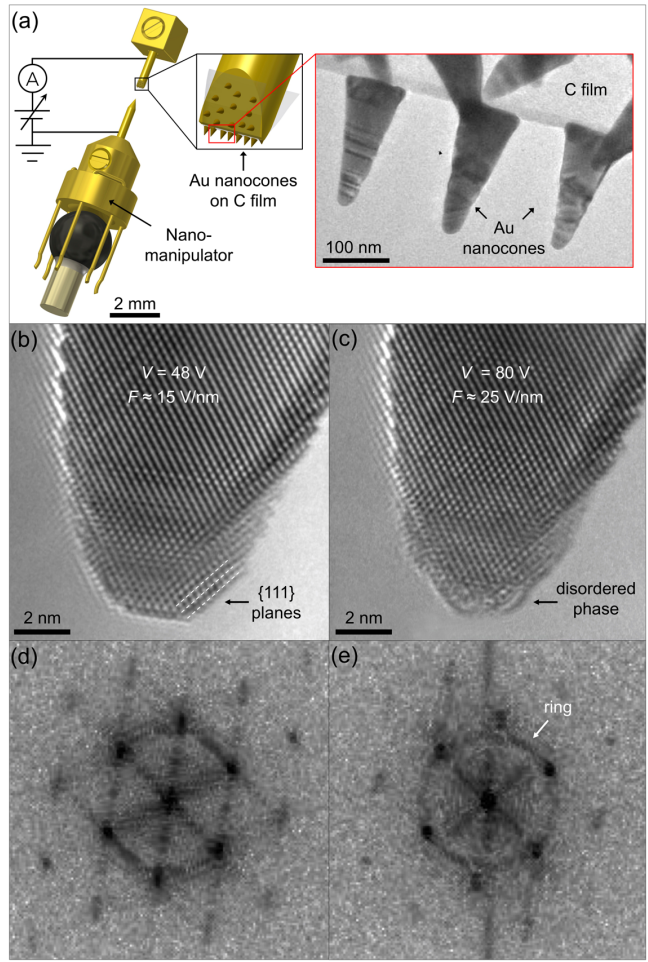

FIG. 1 (color online). Experimental setup and results of electric-field-induced order-disorder switching of atomic layers. (a) Schematic of parts of the *in situ* TEM sample holder with the nanomanipulator and the corresponding electrical circuit. The large inset shows a TEM micrograph of the investigated Au nanocones on a carbon (C) film. (b)–(c) 10 nm of the apex of one of the nanocones as frames extracted from *In Situ* TEM Movie 2 in the Supplemental Material. The disordered state was switched off and on, respectively, by applying a voltage $V$ of (b) 48 V resulting in an electric field $F$ of around 15 V/nm and (c) $V = 80$ V and $F \approx 25$ V/nm. (d)-(e) Fast Fourier transformations of the apex area of (b) and (c). The ring in the FFT in (e) indicates a disordered phase.

In order to rule out that the observed order-disorder transition occurred due to the high mobility of Au atoms at the tip or temperature effects, we compared the mobility of the atoms in the presence and absence of both the electron beam and electric field, and estimated the temperature increase induced by the beam. In particular, the high degree of dangling bonds at a step can increase the Au atom mobility and cause a sharp tip to become shorter and blunter even without exposure to the electron beam [29]. The energy transferred from the incident electron beam can then further increase the mobility. Similarly, the temperature increase from the beam-specimen interaction could possibly affect the crystallinity of the nanocones.

During a 4 min long electron beam exposure without electrical bias, we observed that the tip radius $r$ of the Au nanocone remained constant but that the nanocone was shortened by 5 atomic layers caused by a redistribution of Au atoms by surface diffusion ($t$ = 0-4 min in Fig. 2(a), and corresponding Figs. 2(b) and 2(c) extracted from *In Situ* TEM Movie 1 in the Supplemental Material). After turning off the beam for 28 min [$t$ = 4-32 min in Fig. 2(a)], redistribution of atoms resulted in a shortening of the nanocone by another 5 atomic layers [Fig. 2(d)]. Hence, the electron beam was found to give rise to an increase of the atom diffusion rate by a factor of about 7, but it was not observed to affect the crystallinity of the nanocone [cf. Fig. 2(b) and 2(c)].

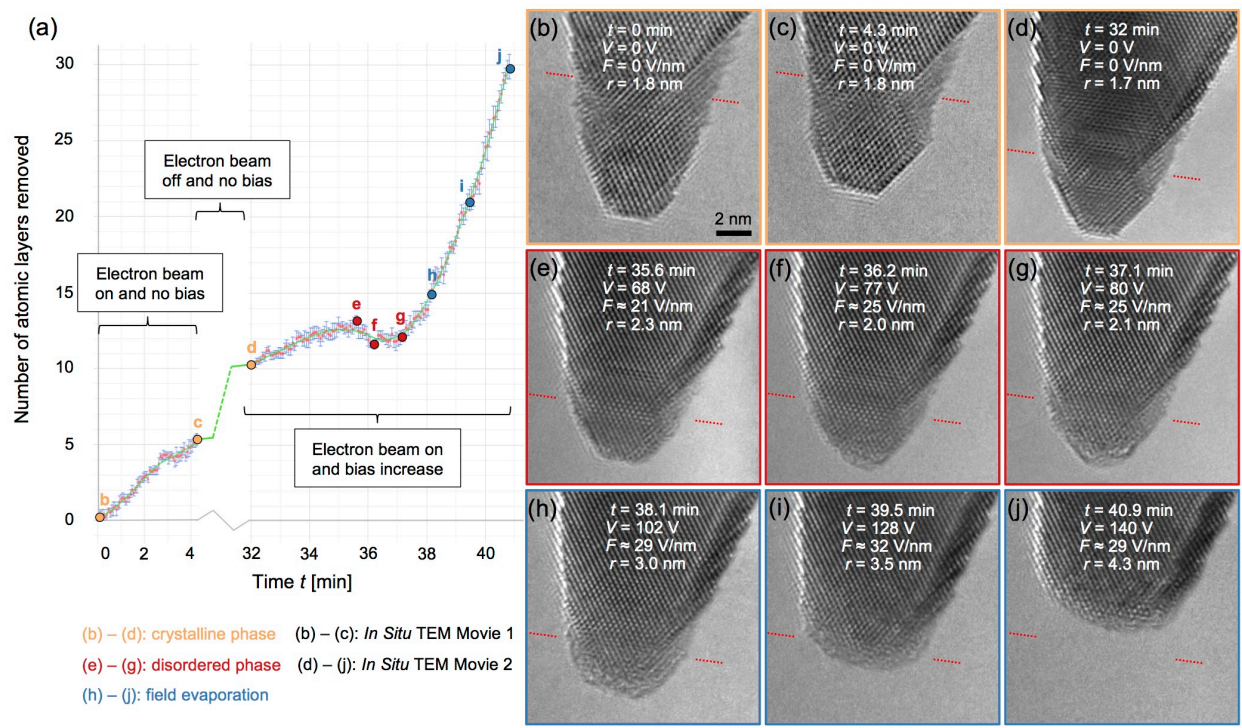

FIG. 2 (color online). Evolution of a Au nanocone apex under an electron beam and an intense electric field; from ordered to disordered to field evaporated phase. (a) Number of atomic layers removed as a function of time $t$. (b)–(j), Extracted frames from *In Situ* TEM Movies 1 and 2 in Supplemental Material, corresponding to the points labeled b – j in (a). In the first and crystalline regime [orange, (b)–(d)], no electrical bias was applied and the electron beam was first turned on, then off and then turned on again (note the truncated $x$-axis); in the second and disordered regime [red, (e)–(g)], the electron beam was on and a bias from 68 to 80 V was applied; in the third regime [blue, (h)–(j)], the electron beam was on and the external bias was increased until field evaporation occurred (102-140 V). The error bars in (a), with an average height of 0.9 atomic layers, represent the uncertainty in measuring the number of atomic layers. $V$, $F$ and $r$ denote the applied voltage, the corresponding electric field and the tip radius, respectively. The red dotted lines in (b)–(j) mark points of reference. The scale bar in (b) also applies to (c)–(j).

The temperature increase that was induced by the electron beam could be estimated through the increase of the atomic diffusion rate using the Arrhenius equation

$$k = A e^{-E_a/RT} \quad (\text{Eq. 1})$$

where $k$ is the diffusion rate, $A$ is the pre-exponential factor, $E_a$ is the activation energy, $R$ is the gas constant and $T$ is the temperature. Using Eq. 1, the sample temperature under the beam can be expressed as

$$T_{on} = \frac{E_a}{R\left(\ln\frac{k_{off}}{k_{on}} + \frac{E_a}{RT_{off}}\right)} = 316.5 \pm 3.5 \text{ K}.$$

with the subscripts "off" indicating the absence, and "on" the presence of the electron beam, and with the Au surface diffusion activation energy $E_a = 21$ kcal/mol [30], $T_{off} = 300$ K and $R = 8.314$ J mol$^{-1}$ K$^{-1}$. Thus, the 300 keV electron beam leads to an increase of the temperature of the nanocone by $16.5 \pm 3.5$ K, taking into account the same uncertainty as in the number of removed atomic layers [the same as the error bars in Fig. 2(a)]. Such a small temperature increase is expected to have a negligible effect on the emergence of a disordered phase and on the order-disorder switching, considering the significantly higher melting temperature (700-800 K) of Au nanoparticles [31].

Consequently, only upon application of an electrical bias to the nanocone, which was increased from 0 to 80 V ($t = 32\text{-}37$ min in Fig. 2(a), and corresponding Figs. 2(d) to 2(g) extracted from *In Situ* TEM Movie 2 in the Supplemental Material), the development of a disordered phase could be observed. It started to grow at an electrical bias of 68 V [Fig. 2(e)], and at 77 V and 80 V (Figs. 2(f) and 2(g), respectively), some of the outmost atomic layers were found to enter a disordered state. Between 68 and 77 V, a slight decrease in the tip radius from 2.3 to 2.0 nm was observed [Figs. 2(e), 2(f) and the dip in the graph at $t = 36$ min in Fig. 2(a)], which is assumed to be owing to the increased mobility of the atoms in the electric field that momentarily sharpens the tip.

Notably, the order-disorder switching was found to be fully reversible using voltage control. The reversibility is seen as a re-crystallization of the disordered layers as the electric field is decreased (Fig. 3, extracted frames from *In Situ* TEM Movie 3 in Supplemental Material). This was achieved on a different Au nanocone as the one shown in Figs. 1(b), 1(c) and 2 (thus also demonstrating the reproducibility of the electric-field-controlled order-disorder transition). In Fig. 3(a) a voltage $V = 78$ V and an electric field $F$ of around 28 V/nm is applied to the nanocone and the outmost atomic layers are in a disordered state. In (b) and 27 s later, $V = 49$ V and $F \approx 17$ V/nm with the disordered state still being present. In (c) and 7 s later, at $V = 40$ V and $F \approx 14$ V/nm, the disordered layers have returned to their original state, from the decrease in the applied voltage. In the beginning of *In Situ* TEM Movie 3 the voltage is kept constant, with the disordered layers remaining unchanged. That is, the disordered layers neither evaporate nor re-crystallize at a constant voltage. Only in the second part of the movie, when the voltage is decreased, do the disordered layers re-crystallize.

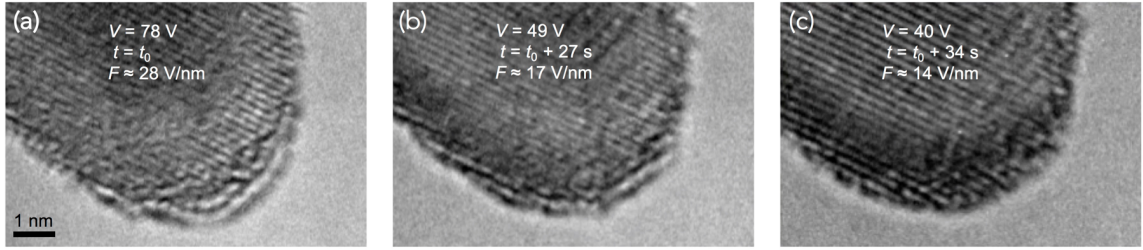

FIG. 3. Reversibility – from disordered to ordered during voltage decrease. The apex of a Au nanocone (different from the nanocone in Figs. 1 and 2) with tip radius of 2.5 nm and distance to cathode of 58 nm at different voltages. (a) The outmost layers are in a disordered state, at an applied bias $V$ of 78 V and an electric field $F$ of around 28 V/nm. (b) 27 s after (a), with $V = 49$ V and $F \approx 17$ V/nm, some of the outmost layers remain in a disordered state and in (c), 7 s later, the nanocone has reverted to a crystalline state, at a bias of 40 V and an electric field of around 14 V/nm.

To distinguish the observed order-disorder switching from field evaporation of the surface layers, we investigated the structure and size of the first nanocone (*i.e.*, not the nanocone in Fig. 3 that we demonstrated reversibility on) also at higher electrical biases of up to 140 V [$t = 38$-$41$ min in Fig. 2(a), and corresponding Figs. 2(h) – 2(j)]. We found that the Au atoms of the disordered state started to field-evaporate at an electrical bias of just below 102 V, corresponding to an electric field of around 29 V/nm [Fig. 2(h)]. Overall, the number of atomic layers removed increased sharply in the field evaporation regime [$t = 38$-$41$ min in Fig. 2(a)] compared to lower or no electrical biases. During this period, the nanocone experienced a shortening with around 18 atomic layers, and a broadening with an increase in the tip radius from 3.0 to 4.3 nm [Figs. 2(h) – 2(j)]. Although the disordered layers evaporated, new disordered layers were immediately formed from the next atomic layer in the crystal. As a clarification, evaporation occurred when increasing the bias, whereas the re-crystallization mentioned before occurred when the bias was decreased.

In order to obtain deeper understanding of the mechanism and temperature-dependence of the observed electric field-induced order-disorder transition, we conducted a series of atomic scale simulations based on density functional theory (see Methods in the Supplemental Material for details). This approach accounts for charge redistribution and screening at the surface at a quantum mechanical level. The objective is to be as predictive as possible and minimize ambiguities due to approximations intrinsic to resort to semi-empirical models [32,33].

As a first step, we determined the field evaporation threshold from *ab initio* molecular dynamics (AIMD) simulations. To this end, we followed the evolution of an atomistic tip model composed of 192 atoms at different constant external electric fields at 300 K. The AIMD simulations revealed an onset of evaporation at a field strength of around 25 V/nm, which is in good agreement with the experimental observations (see AIMD Movie 3 in the Supplemental Material). The simulations, however, failed to show disordering, which is not surprising given the constraint on the atomic geometry due to the fixed bottom layer in these simulations (Discussion 5 in the Supplemental Material).

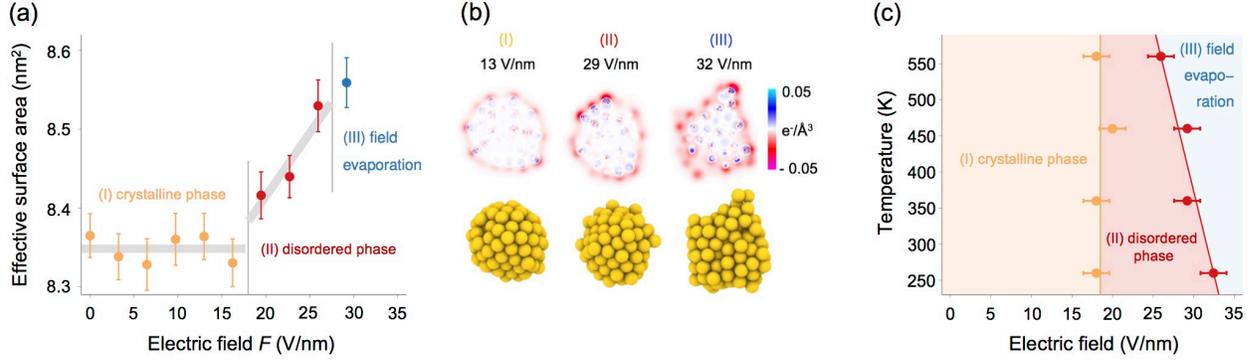

FIG. 4 (color online). *Ab initio* simulations of surface roughening. (a) Effective surface area of a gold nanoparticle at 360 K as a function of the electric field. (b) Excess charge distribution (top) and shape (bottom) for three gold nanoparticles representative of the three stages indicated in (a). The color scale indicates the accumulation (light blue, maximum of 0.05 electrons/Å$^3$) and depletion (cerise, minimum of -0.05 electrons/Å$^3$) of electronic charge. (c) Approximate phase diagram showing the temperature dependence of the electric fields at which disordering (II) and field evaporation (III) occurs.

Given the experimental observation of surface disorder at large fields, one might suspect that the electric field causes a softening of the bonds at the surface, leading to much larger thermal displacements. This would effectively suggest a form of surface melting in analogy to the Lindemann criterion, which formulates an empirical relation between the melting point and the mean square displacement of the atoms [13]. To test this hypothesis, we investigated the relaxation, charge distribution and atomic thermal displacements at planar {111} Au surfaces under the influence of an external electric field (Fig. S4). The large density of states at the Fermi energy gives rise to a strong dielectric response, which very effectively screens the external electric field. Therefore, even for the largest field strengths of over 30 V/nm considered here, the external field barely penetrates more than one or two atomic layers [Fig. S4(d)]. While the top layer is strongly affected by the field, already the third – and for smaller field strengths even the second layer – exhibits bulk behavior. The localization of the excess charge at the topmost surface layer causes an outward relaxation of about 1% at a field strength of 30 V/nm [Fig. S4(a)]. The thermal displacements are indeed enhanced for the first layer and at a field of 26 V/nm, the thermal displacements at 300 K are 25% larger than in the field-free case. This is, however, only equivalent to a temperature increase of about 100 K under field-free conditions and thus insufficient to explain the experimental observation [Fig. S4(b)].

To overcome the limitations imposed by the geometrical constraints intrinsic to tip model and extended surfaces (Discussion 5 in the Supplemental Material), we considered the behavior of small Au nanoparticles containing 135 atoms as a function of temperature and field strength. This allowed us to study the surface evolution not only without geometric constraints, but simultaneously to explore a reasonably wide range of temperatures and field strengths. The simulations showed that the electric field induced a very pronounced atomically-localized charging of the surface [Fig. 4(b)]. At low to modest field strengths, the particles retained a spherical shape with an approximately constant surface area (stage (I) in Fig. 4 and AIMD Movie 1 in the Supplemental Material). If the field strength exceeded 18 V/nm, the particles elongated and displayed pronounced surface modes leading to a steady increase of the surface area with increasing field strength (stage (II) in Fig. 4). This change was accompanied by surface disordering and the localization of excess charge at individual surface atoms protruding from the surface [top row in Fig. 4(b)]. The shape elongation along with the occurrence of individual atoms at the surface support the hypothesis that the observed transition can be understood as a form of surface roughening, making it more energetically favorable to form surface defects. Ultimately, if the field strength was increased further, field evaporation took place (stage (III) in Fig. 4, AIMD Movie 2) as observed experimentally.

Although phase transitions in few-atom systems are often difficult to detect, here, the pronounced variation of the surface area with electric field provides a clear indication. Using the change in surface area [Fig. 4(a)], we can extract the roughening (disordering) onset field and the field evaporation threshold as a function of temperature [Fig. 4(c)]. While the roughening transition appears to be relatively insensitive to temperature at 18 V/nm, the field evaporation threshold is found to decrease from 33 to 28 V/nm between 260 and 560 K, which is in semi-quantitative agreement with previous experimental data [34]. Given the simplifications intrinsic to the simulations, in particular the use of nanoparticles, the agreement of the onset fields for disordering and evaporation with the experiments is remarkable. In all, these findings suggest a generality of electric-field-induced surface disorder also for other types of nanostructures and metals.

In summary, we have reported an electric field-controlled and reversible order-disorder switching mechanism of the topmost atomic layers of Au nanocones, which was experimentally observed with atomic resolution using *in situ* TEM and confirmed by *ab initio* simulations. The switching is contactless, which has several advantages over no contact-free switching, e.g. mechanical wear and stress on a contact. The switching occurs at high electric fields that are difficult to reach using planar surfaces. Typically, nanometer-sized apices greatly enhance the electric fields [22], as for the Au nanocones. To extend our results for larger regions, devices with varying length scales can be nanostructured using existing technologies for electronics and sensing. The findings suggest that the order-disorder transition can be seen as electric field-induced surface roughening, providing fundamental understanding of atomic structural dynamics. Even though atom probe tomography operates at low temperatures [35], such knowledge is crucial for this and other nanomaterial characterization techniques dependent on intense electric fields (several large projects are aiming to combine atom probe tomography and TEM [36]). Of an immediate relevance is the external dynamic control of surface processes such as light-matter coupling via localized surfaces plasmons, which in noble metals typically are localized at a few atomic surface layers. With this work we uncover a new technique for potentially steering the emergence, propagation and routing of such surface-bound optical modes. This would have broad implications for nanophotonics, nano-photocatalysis and other light-driven processes at the nanoscale. Moreover, the deterministic manipulation of the crystallinity of surface layers in nanostructures could be utilized in novel active nanodevices for various applications in field-effect transistor technology, catalysis and sensor technology, as well as in studies addressing fundamental aspects in surface physics, material science and low-dimensional phases of matter.

L.D.K. and E.O. designed and conceived the experiment. L.D.K. carried out the experiment and COMSOL simulations. M.J.K and J.L performed the MD and DFT simulations. K.L fabricated the sample. L.D.K. prepared the manuscript. All authors discussed the results and contributed to writing the manuscript.

This work was supported by the European Network for Electron Microscopy (ESTEEM2, European Union Seventh Framework Program under Grant Agreement 312483-ESTEEM2 (Integrated Infrastructure Initiative–I3)), the Knut and Alice Wallenberg foundation, the Swedish Research Council and the Area of Advance Nano.

# Supplemental Material for
## *Electric field-controlled reversible order-disorder switching of a metal tip surface*

*Ludvig de Knoop[1,‡]\*, Mikael Juhani Kuisma[1,2,‡], Joakim Löfgren[1], Kristof Lodewijks[1], Mattias Thuvander[1], Paul Erhart[1]\*, Alexandre Dmitriev[1,3,4]\*, Eva Olsson[1]\*.*

[1] Department of Physics, Chalmers University of Technology, Gothenburg, Sweden
[2] Department of Chemistry, University of Jyväskylä, Jyväskylä, Finland
[3] Department of Physics, University of Gothenburg, 412 96 Gothenburg, Sweden
[4] Geballe Laboratory for Advanced Materials, Stanford University, Stanford, California 94305-4045 USA

The Supplemental Material document includes 1. Movie Captions, 2. Methods (experimental and computational details), 3. Supplemental Discussions 1-5 and 4. Supplemental Figures S1-S4.

## 1. Movie Captions

*In Situ* TEM Movie 1. A 4 min 16 s long *in situ* TEM movie that shows how the Au nanocone reacts under the 300 keV electron beam. The first and the last frame of the video is found in Fig. 2(b) and 2(c), respectively. No bias was applied to the nanocone. The video is compressed but otherwise unprocessed.

*In Situ* TEM Movie 2. Originally a 8 min 54 s long *in situ* TEM movie that shows how the Au nanocone reacts under the 300 kV electron beam while an applied bias is increased from 0 to 140 V. The video is processed to center the movie around the nanocone by translating each frame along $x$ and $y$ directions and using a denoising software (as described in Methods in the Supplemental Material). Duplicate and blurry frames have been removed. The movie is sped up around 9 times. Frames displayed in Figs. 1(b), 1(c) and 2(d) – 2(j) are extracted from this movie.

*In Situ* TEM Movie 3. Originally a 1 min 59 s long *in situ* TEM movie that shows how the outmost disordered atomic layers of a Au nanocone re-crystallize when the bias is decreased. During the first 50 s the applied voltage is kept constant at $V = 89.6$ V, showing no change to the thickness of the disordered area, but at voltage decrease, the disordered layers re-crystallize. The denoising software described in Methods in the Supplemental Material has been used. The movie is sped up 4 times. Annotations display the externally applied voltage $V$ and the time $t$. Frames displayed in Figs. 3 are extracted from this movie. The Au nanocone is different from the nanocone in *In Situ* TEM Movie 1 and 2.

AIMD Movie 1. *Ab initio* molecular dynamics simulations ($T = 360$ K) for a gold nanoparticle at different field strengths. The surface mesh is calculated with a probe sphere radius of 3 Å and smoothing level 4 (Ovito parameters). The coordination analysis is done by counting the number of neighboring atoms within a radius of 3.4 Å.

AIMD Movie 2. *Ab initio* molecular dynamics simulations of gold nanoparticles at different temperatures (260, 360, 460 and 560 K) and different electric field strengths. The movie illustrates the rapid evaporation in the case of particles with larger surface charge. After a field evaporated atom is too close to the unit cell boundary, it is removed assuming a charge of +1. The runs after field evaporation events are used here only for visualization of the field evaporation.

AIMD Movie 3. *Ab initio* molecular dynamics simulations ($T = 300$ K) for gold nanotip model in different external electric fields. The bottom layer is constrained to preserve the geometry.

## 2. Methods

***In situ* TEM experiment.** For the experiment, an *in situ* transmission electron microscopy (TEM) sample holder (Nanofactory Instruments) was used. It can apply a bias of ±140 V and is fitted with a nanomanipulator. The holder was used inside a Titan 80-300 TEM (FEI Co.), which was equipped with (for this experiment non-excited) monochromator, Cs probe-corrector and Gatan image filter. The acceleration voltage of the TEM was set to 300 kV. During the experiment depicted in Figs. 1 and 2, the illumination system of the TEM generated a current on the fluorescent screen of 2.5 nA. Before the measurements, the objective astigmatism was corrected for by using a live fast Fourier transformation (FFT) window in Digital Micrograph. By defocusing on an amorphous part of the sample, a set of ovals appeared in the live FFT window. The objective astigmatism was adjusted until the ovals turned into rings, which verified that the astigmatism was at a minimum. Then, the sample was set to be in focus, which was verified in the live FFT window by disappearance of the rings. Distance-calibration was achieved using the spacing between the {200}-planes [see Fig. S2(l)].

The piezo-driven nanomanipulator [left-hand part of Fig. 1(a)] of the sample holder could move in three dimensions with a range of 2 mm and a spatial resolution of 0.02 nm to 0.2 nm (depending on direction of movement). The nanomanipulator utilizes a slip-stick motion by moving a piezo [the part below the black sphere in Fig. 1(a)] slowly in one direction (stick) and fast in the opposite direction (slip) [1]. As probe, a Au tip was inserted into the nanomanipulator. The tip was fabricated from electrochemical etching using 37% HCl with a Pt electrode and the end part of the Au tip covered in Lacomit. 30 s after the etching process was terminated the tip was dipped once in deionized $H_2O$ and once in ethanol.

The Au nanocones were produced with hole-mask colloidal lithography [2]. The cones are arranged on an electron-transparent carbon (C) film in a macroscopic short-ranged-ordered array [3]. They typically feature a tip radius $r \geq 5$ nm and a height of around 180 nm. The C film with nanocones was transferred to a mechanically cut Au wire, with the C film wrapped around the Au wire. Subsequently, the Au wire with nanocones was inserted into the *in situ* TEM sample holder [Fig. 1(a)].

Using the nanomanipulator of the sample holder, one of the Au nanocones from the C film was transferred to achieve direct contact with the Au wire of the sample, in order to ensure a good electrical contact. The Au tip on the nanomanipulator was flattened out via a controlled explosion; this was accomplished by increasing the voltage and short-circuiting the electrical circuit by going into contact with the Au wire that the sample was wrapped around. The flat Au tip was connected to ground, *i.e.*, it was the cathode.

The nanocone of Fig. 2(b) exhibited at the beginning of the experiment a tip radius *r* of 1.7 nm. Initially, when attached to the C film, the tip radius was around 5 nm. To reach a radius of 1.7 nm, the cone was brought into contact with a small protrusion on the flattened Au tip. By increasing the bias to 320 mV, a current of 550 µA was generated that plastically deformed the apex of the cone. In this stage, the nanomanipulator was retracted at a speed of 0.7 nm/s until the cone was separated from the protrusion on the flattened Au tip, forming the 1.7 nm tip radius seen in Fig. 2(b). The current density *J* at the end of the nanocone at the moment of separation was $J \geq 6.1 \times 10^{13}$ A/m$^2$. Similar tip-sharpening techniques have been employed before [4,5]. Finally, using the nanomanipulator, the flat part of the Au tip was positioned opposite and a distance of 100 nm from the Au nanocone.

Considering the close proximity of the Au nanocone and the flat cathode (100 nm), an electric field in the order of 20-30 V/nm extorts an attractive force between the two parts large enough to close the distance to the cathode, resulting in a destroyed nanocone and cathode. For anyone aiming to reproduce the results, it is recommended to be ready to either lower the bias or increase the separation distance at the first sign of the nanocone (or similar) starting to move towards the cathode.

**Video-analysis.** The electric field created around the nanocone apex by the externally applied potential affected the TEM electron beam, which was observed as a lateral movement of the beam. To move the image back into view, the image shift functionality of the TEM was used. The original biasing movie, recorded at around three frames per second with a screen capture software (CamStudio), thus displays the Au nanocone moving back and forth. In order to facilitate observation of changes on the atomistic level of the nanocone, the movie was centered by translating each frame in *x* and *y* (using the add-on C2 TrackX (CoreMelt Pty. Ltd.) to the program Final Cut Pro 10.1.3 (Apple Inc.)). To reduce white noise in the movies, a moving averaging over 3 frames was performed using the Final Cut Pro add-on Neat Video (Reduce Noise v4.5.0) (*In Situ* TEM Movie 2 in the Supplemental Material). The following settings were employed for the denoising add-on: In the Temporal tab, the Temporal Filter was enabled and settings were Quality Mode = High, Radius = 1, Noise Level = -60% and Amount = 100%. Adapt to Changing Noise was enabled. In the Spatial tab, the Spatial Filter was disabled. In the General tab, Mix with Original was disabled. The TEM images in Figs. 1(b) – (c) and 2(d) – (j) are frames extracted from *In Situ* TEM Movie 2, which was treated in this way. To show that no additional features were added by using a noise-reduction software, the difference between treated and untreated frames can be seen in Fig. S1. The white noise reduction method mentioned above was also utilized on *In Situ* TEM Movie 3. The *In Situ* TEM Movie 1 is unprocessed.

**Number of layers removed.** The number of removed layers that are displayed in Fig. 2(a) was calculated by measuring the total length of removed material along the nanocone axis divided by a measured inter-atomic-plane distance. The second movie (*In Situ* TEM Movie 2 in the Supplemental Material) contained 1654 frames (after having removed duplicate frames), obtained during 8 min and 54 s. During the experiment, occasionally fast vibrations of the nanocone resulted in a total of 197 blurry frames, which were discarded. Of the remaining 1457 frames, every 11th (sometimes every 12th) frame was used to quantify the morphological evolution of the nanocone. For each of these 131 frames, the maximum and minimum number of the atomic layers that were removed were measured. For each data point, two measurements were made and a median value was calculated. The difference between the highest and lowest value provides the error bars displayed in Fig. 2(a). The red dots in Fig. 2(a) correspond to median values of the highest and lowest values. The first part of the graph in Fig. 2(a) (labeled "Electron beam on and no bias"), corresponds to data from the first movie (*In Situ* TEM Movie 1 in the Supplemental Material), which is unprocessed. The curve was based on the red dots in the graph using the smoother function of the software Jmp12 (lambda value of the smoother set to 0.015). The nine orange, red and blue filled circles indicate the particular frame from the movies displayed in Figs. 2(b) – 2(j). The time between each measurement point in Fig. 2(a) was 4.1 s.

**Tip radius.** The tip radii measurements in Figs. 2(b) – 2(j) were obtained by taking the average value of two measurements.

**Inverted $\log_{10}$ of the FFT of the Hanning-window of an average of 100 frames.** Figs. 1(d) and 1(e) and Figs. S2(c), S2(f), S2(i) and S2(l) were obtained by inverting an image that was created taking the $\log_{10}$-value of a fast Fourier transformation (FFT) of a Hanning-window of an average of 100 frames:

$$1 / \left[ \log_{10} \left( FFT \left\{ W_H \left( \frac{1}{n} \sum_{i=1}^{n} frame_i \right) \right\} \right) \right], n = 100$$

where $W_H$ is the Hanning-window function that is applied to the selected area (which has the dimensions of an integral power of two). The Hanning-window script that was used; "hanning_FFT.s" was created by Ruben Bjorge (v.1.0 2007-10-09). The averaging was done using Digital Micrograph (Gatan Inc., www.gatan.com). That is, the frames were *not* denoised with the method described above under Video-analysis.

**Finite Element Method simulations.** To estimate the electric field of the Au nanocone at room temperature, finite element method (FEM) modeling was employed (COMSOL Multiphysics software, www.comsol.com). A 3D model was created from a 2D axisymmetric model of the Au nanocone by revolving the simulation 180° around the center vertical axis of the Au nanocone simulated electric field maps (see Fig. S3). The *Electrostatics* model and the default value of 4.56 $\times 10^7$ S/m for the electrical conductivity of Au were used. The electric field is displayed using the "Surface Electric Field norm", meaning that the expression *es.NormE* was used, which is defined as $es.NormE = \sqrt{\left|F_x\right|^2 + \left|F_y\right|^2 + \left|F_z\right|^2}$, where $F$ is the electric field in the *x*, *y* and *z* direction, respectively. The dependence of the electric field on atomistic variations is shown in Fig. S3.

**Atomistic *ab initio* molecular dynamics simulations with density functional theory**
I. COMPUTATIONAL DETAILS
A. Surface relaxation and energetics
   The surface relaxation was analyzed for the {111} surface by means of density functional theory (DFT) calculations. Slab models comprising up to 20 layers were considered. The plane wave mode of GPAW was used with a plane wave energy cut-off of 340 eV while the Brillouin zone was sampled using a Γ-centered $12 \times 12 \times 1$ *k*-point mesh [6]. The external electric field was applied as described in Sect. II A below, while the energetics were analyzed as detailed in Sect. II B.

B. Thermal vibrations

To analyze the thermal displacements at the {111} surface finite displacements calculations were carried out, from which the force constant matrix was constructed. These calculations were prepared and analyzed using the PHONOPY package [7]. In these calculations, we employed slab models comprising 9 atomic layers with a cross section of 3 × 3 surface unit cells. The plane wave mode of GPAW was used with a plane wave energy cutoff of 340 eV while the Brillouin zone was sampled using a Γ-centered $4 \times 4 \times 1$ $k$-point mesh [6].

C. Molecular dynamics simulations

DFT calculations were performed using a basis set obtained by linear combination of atomic orbitals [8] as implemented in the projector augmented wave method code GPAW [6]. Using tools in the atomic simulation environment [9], we ran *ab initio* Born-Oppenheimer molecular dynamics (MD) simulations with a time step of 15 fs. The canonical ensemble was emulated using a Langevin thermostat. The grid spacing was 0.22 Å and a custom double zeta basis set was used throughout. The basis set was constructed similarly as the silver basis set in Ref. 10. A similar computational setup using a larger time step has been previously found to work well for coarse sampling of gold clusters [11]. We note that hybrid classical-electrodynamical methods have been developed to model field evaporation [12,13]. Here, we rather adopt a fully quantum-mechanical approach and accept a compromise with respect to the attainable system sizes. This choice is motivated by a desire to be as predictive as possible and minimize ambiguities due to approximations intrinsic to semi-empirical methods.

*1. Simulations of model tip*

The objective of the first set of MD simulations was to reproduce the experimental conditions as closely as possible, specifically the tip shape, and to describe the process of field evaporation. The purpose of these MD simulations was primarily to provide *qualitative* insight into the process of field evaporation and demonstrate that the field strength at which evaporation occurs is in reasonable agreement with experiment.

A model tip was constructed by cutting out a truncated cone from a bulk Au crystal, which was subsequently relaxed (in the absence of an electric field) via MD simulations using the LAMMPS code [14] and an embedded atom method potential [15]. The obtained structure containing 192 atoms was subsequently used as starting point for MD simulations within the DFT framework as described above. In these simulations, a static electric field was imposed, which was linearly increased during the course of the simulation. Also, note that the system size, which was dictated by computational cost, was too small to properly capture surface disordering, as the latter involved a number of surface layers that was comparable to the total system size, and time scales that exceed the ones accessible here.

*2. Nanoparticle simulations*

In addition to the model tip, we considered a nanoparticle comprising 135 atoms. Nanoparticles of this size can be expected to have a lower free energy barrier for melting than surface slabs or the model tip (which includes a constrained bottom layer). They are therefore better suited for our sampling structural transitions in the presence of an electric field. Furthermore, we assert that due to metallic screening, the electric field effect is localized at the surface and hence the exact geometry is of lesser importance for the phenomenon. This also implies that the surface charge density can act as a macroscopic parameter of the system. We furthermore note that a finite cluster may theoretically have a finite band gap at full shell occupation. However, removing a charge is not a linear response effect but a fully self-consistent renormalization of orbitals and metallic behavior can be observed at all charge states, *i.e.*, all charge accumulated at the surface.

In the case of these nanoparticles, the effect of the external electric field was captured by removing electrons from the system. This approach avoided ambiguities related to the orientation of the electric field and has been extensively verified by comparison to slab calculations.

We performed an array of MD simulations sampling 15 different electric field strengths, by removing up to 22.5$e$ from the system, and four different temperatures (260, 360, 460 and 560 K). The effective field strength was calculated using the nominal spherical radius $\frac{1}{4}Na_0^3 = \frac{4}{3}\pi R^3$ with $N$ = 135 and $a_0$ = 4.08 Å. In this fashion, one obtains for example for a charge of $C$ = 16.5$e$, corresponding to a surface charge of $\sigma = C/(4\pi R^2)$, a field strength of $\varepsilon = \sigma/\varepsilon_0 = 35$ V/nm.

Simulations were started from ideal face-centered cubic lattice structures with {100}, {110} and {111} surfaces. As expected for small nano-clusters, all simulations led to configurations, in which the center atom assumed icosahedral symmetry. This demonstrates that the nanoparticles achieved a reasonable sampling of the configuration space.

A total of over 130,000 MD steps was simulated for 60 systems (4 temperatures, 15 different values for the electric field strength), corresponding to a total simulation time of over 2 ns. The surface area of the nano clusters during the MD runs was determined by constructing the convex hull and using Delaunay triangulation as described in Ref. 16. Since the construction of the convex hull was based on the atomic positions, implying that the atoms were treated as point objects (zero volume), the surface area was systematically underestimated. To correct for this effect, the surface area was rescaled by a constant factor to normalize the surface area to the volume per atom.

II. METHODOLOGICAL ASPECTS

A. Modeling surfaces in strong electric fields

The commonly adopted approach to model surface slabs in electric fields is to impose a saw tooth potential by inserting a dipole layer in the vacuum region [17]. This implies the creation of both an anode and a cathode in the same simulation. As a result, the system becomes asymmetric and the surface energetics correspond to a mixture of positive and negatively charged surfaces. Furthermore, the maximum fields that can be achieved are limited by the Fermi energy reaching the vacuum level on the cathode side (equivalent to electron cold-field emission). Using common parameters this implies a maximum field strength of only about 2-5 V/nm, considerably below the range, in which surface disorder and field evaporation is observed experimentally. This issue actually mirrored the experiment, where an optimum setup had to be found in order to achieve a field at the cathode that is smaller than the onset for electron cold-field emission (around 2 V/nm), while simultaneously having a large enough field around the anode nanocone in order to initiate the disordered phase (around 25 V/nm; Discussion 1 in the Supplemental Material).

To circumvent this problem, we effectively imposed a V-shaped potential profile by inserting a monopole layer in the center of the vacuum region, which was compensated by removal of the equivalent amount of charge from the slab, as already described in Ref. 18. In this fashion, we obtained a symmetric system [see Fig. S4(c)], which enabled us to analyze flat surfaces in strong electric fields, removed the limit on the maximum field strength inherent to the dipole layer approach, and furthermore considerably improved the convergence of the self-consistent field loop. In the range of field strengths that can be achieved by the traditional saw tooth approach, the present approach yielded identical results.

In practice, we were interested only in the anode side of the surface and thus focused exclusively on positively charged surfaces. Note, however, that the present approach is equally applicable for studies of the cathode side.

B. Induced surface charge

At a given geometry R, the induced charge can be calculated as $n_{ind}(\mathbf{r}) = n[\mathbf{R};+C](\mathbf{r}) - n[\mathbf{R};0](\mathbf{r})$. For slab calculations, it was convenient to average the induced charge in planes perpendicular to the surface.

## 3. Supplemental Discussions

**Discussion 1: Finding optimum cone tip-radius and separation distance to reach the onset electric field.** Finite element method modeling (FEM) was used to estimate the electric field using methods developed and reported elsewhere [19]. As mentioned in the main text, the *in situ* TEM sample holder can apply 140 V. The electric field $F$ is proportional to the applied potential $V$ and inversely proportional to the distance $d$ that separates the electrodes. If one of the electrodes is shaped as a tip with a tip radius $r$, the field $F$ around the Au nanocone can be described as

$$F \propto \frac{V}{dr}$$

Due to the 140 V limitation of the holder, either the separation distance $d$ or tip radius $r$ must be decreased in order to increase the electric field strength. With a sub-Å control of the nanomanipulator of the sample holder [Fig. 1(a)], the distance $d$ between the Au nanocone (anode) and Au tip (cathode) can be accurately controlled to reach a minuscule separation distance. This approach was tried but it caused a problem. During a similar experiment, not reported here, using $d = 14$ nm and a Au nanocone with $r = 10$ nm, the field around the flat Au tip (cathode) reached the onset value for electron cold-field emission (around 2 V/nm) before the field around the Au nanocone (anode) to switch the surface layer to the disordered state was reached (around 25 V/nm). Simulations using FEM confirmed this. To be certain that any structural changes are not coming from electrons bombarding the anode nanocone, the electric field around the cathode had to be lower than the onset electron field emission field in the order of 1-2 V/nm [19,20]. That is, the distance $d$ between the Au nanocone and the flat Au tip had to be larger. To summarize, we needed a system which produces a field in the order of 25 V/nm around the nanocone anode, while at the same time keeps the field around the flat cathode low enough ($\lesssim 1$ V/nm) to inhibit electron cold-field emission. Further simulations pointed towards a separation distance of $d = 100$ nm and a tip radius of the Au nanocone of $r < 4$ nm to accomplish this. How this was accomplished is covered in Methods.

**Discussion 2: Field-evaporation field lower than that obtained in atom probe tomography.** In atom probe tomography the field to evaporate Au is about 35 V/nm [21]. Note though that this is at low temperature (20-120 K), which can lower the field-evaporation field (e.g. for W the field is lowered by 15 % when going from room temperature to 100 K [22]).

**Discussion 3: Sputtering from the 300 keV electrons.** Possibly, the atoms in the Au nanocone are displaced by the 300 keV electron beam. According to literature, the threshold energy for bulk Au atom displacement is 1320 keV [23,24], but, for atoms at the surface the sputtering energy threshold is lower, being around 380 keV [23]. Consequently, in addition to slightly increasing the temperature in the specimen as discussed in the main text, the incident electron beam could potentially sputter away atoms at the surface perpendicular to the electron beam. However, the disordered layers observed here are mainly in a plane parallel to the beam.

**Discussion 4: Atomistic finite element method modeling simulations of the Au nanocone apex.** Even though the FEM incorporated atomistic-sized structures (Fig. S3), the exact and time-resolved structures on an atomistic level, which can have an impact on the local electric field, are unknown. For example, in Figs. S3(a) – S3(c) simulations of a tip with same tip radius $r = 2.3$ nm, separation distance $d = 100$ nm and applied voltage $V = 68$ V but different atomistic structures, show a variation in simulated electric field varied of 23 ± 6 V/nm. The smooth surface in Fig. S3(a) provided the smallest field of 16 V/nm and the atom-sized sphere added to an atomistic-mimicked surface in Fig. S3(c) gave the largest field of 29 V/nm. Simulations showing large variations of the electric field on an atom probe tip surface has been studied extensively by e.g. Vurpillot et al. [25,26].

**Discussion 5: Geometric constraints in tip simulations.** It should be noted that because of the small system size in combination with the constraint that the atoms in the basal plane of the truncated tip were immobile, effectively prevent us from using the tip model itself for studying surface roughening. The experimental data suggests that 2-3 atomic layers are affected by disordering, which amounts to almost the entire system size in the tip simulations.

# 4. Supplemental Figures

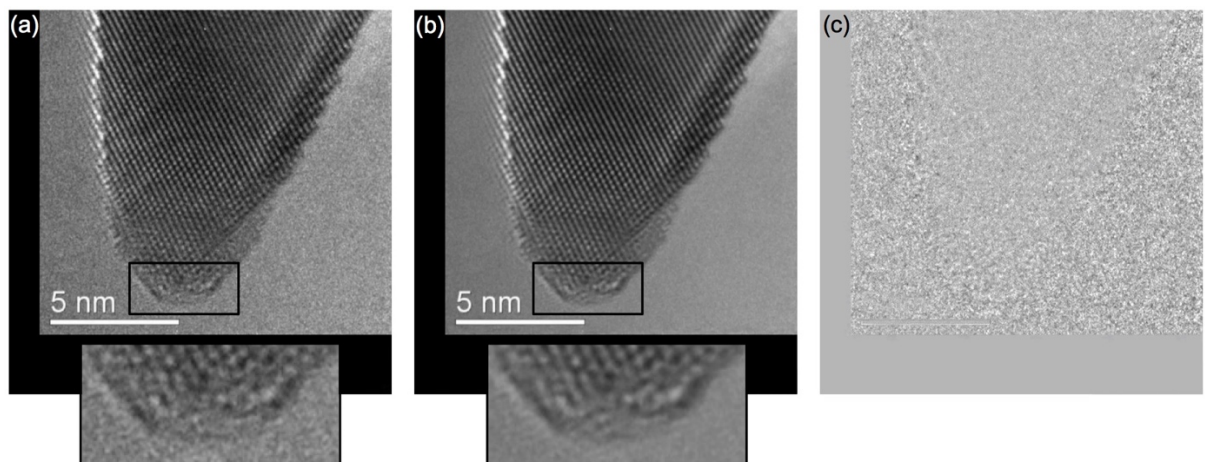

FIG. S1. Comparison of an untreated and corresponding denoised frame. To rule out that any feature was added by the denoising software, a comparison between (a) an untreated frame and (b) the corresponding denoised one was made. The insets in (a) and (b) are zoom-ins of the apex of the Au nanocone. In (c), the result of a subtraction of (b) from (a) is displayed. The frame in (b) is the frame in Fig. 2(g) of the main text.

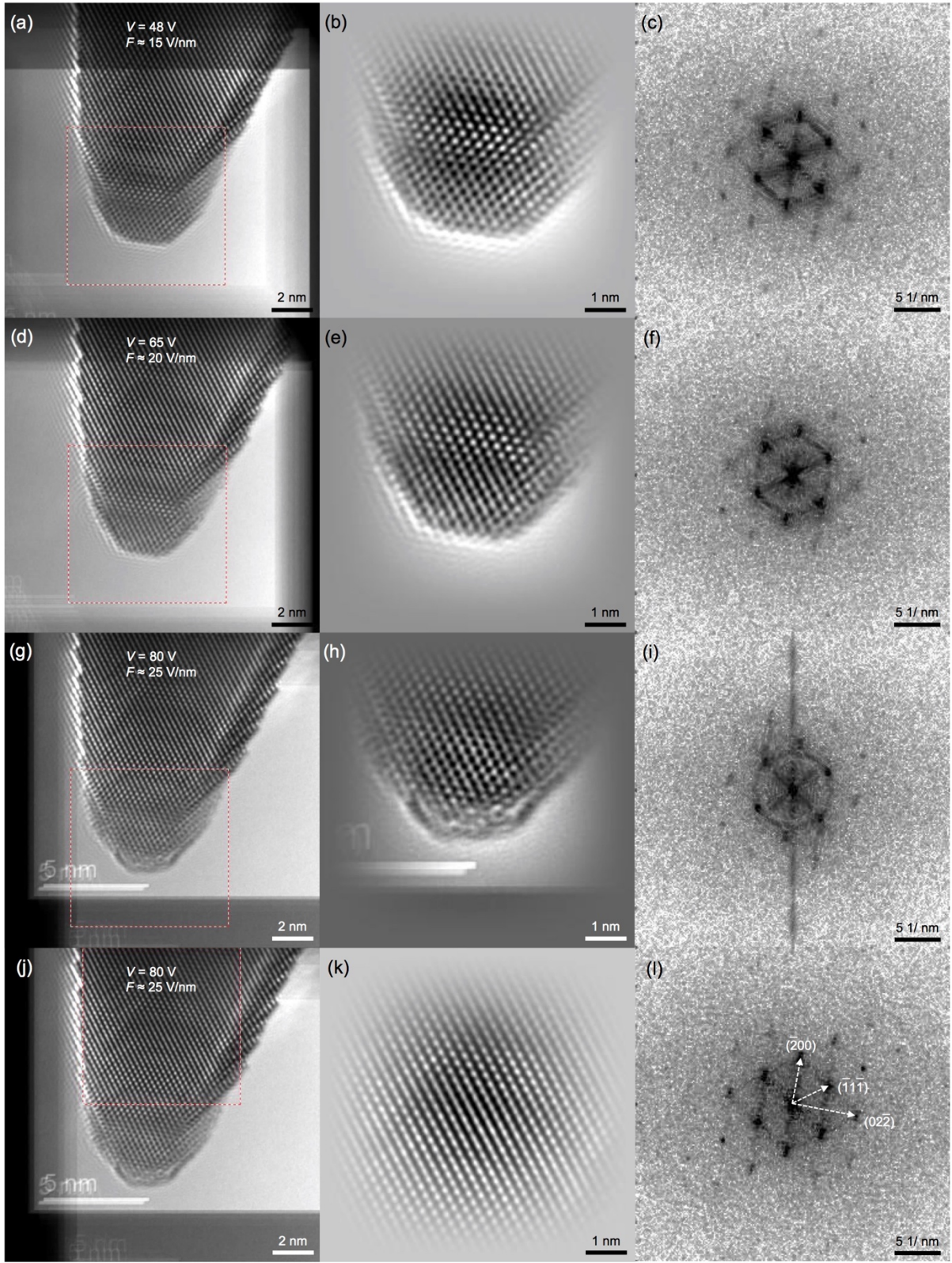

FIG. S2. Method of extraction of fast Fourier transformations from the Au nanocone. Images (a), (d), (g) and (j) are averages of 100 frames at different voltages. Annotated on each image are the voltage $V$ and corresponding electric field $F$ from finite element method simulations. The dotted red squares show the areas where the Hanning-window script was applied (Digital Micrograph script 'hanning_FFT.s' by Ruben Bjorge (v.1.0 2007-10-09)). The results are shown in images (b), (e), (h) and (k), respectively. The fast Fourier transformations of images (b), (e), (h) and (k) can be seen in images (c), (f), (i) and (l), respectively. In (i), a ring can clearly be seen, which is indicative of a disordered phase. Note that the FFT in (l), obtained further away from the apex of the nanocone, shows no ring but a diffraction pattern typical for a bulk crystalline metal. The diffraction pattern was indexed in (l) and corresponded to an incident electron beam along the [001] direction. The ring in (i) coincides with the {111}-spots. The {111}-spots correspond to a lattice spacing of 2.36 Å and the ring radius in (i) is measured to 2.4 ± 0.5 Å. The experimental value reported for melted Au is at a distance of 2.38 Å [27].

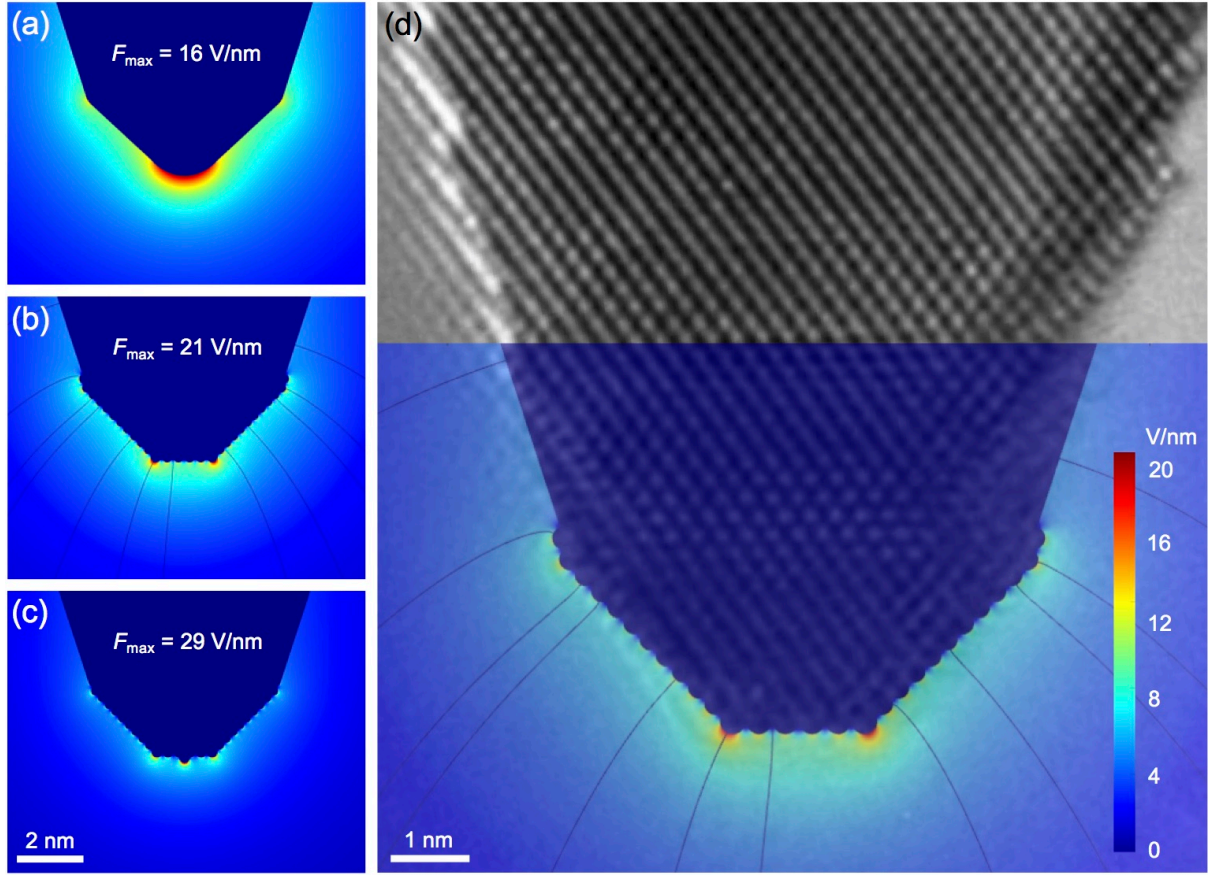

FIG. S3. Dependence of simulated electric field due to atomistic variations of the Au nanocone. Changes on the atomic level could change the electric field, as is shown here. In (a), (b) and (c) slices through the center of the 3D finite element method (FEM) model are displayed (see Methods). The smooth surface in (a) gives a maximum electric field $F_{max}$ = 16 V/nm. In (b), the simulations feature the apex surface covered by atom-sized semi-spheres, resulting in an electric field of 21 V/nm. By adding an atom-sized sphere between the two center semi-spheres from (b), the field increases to 29 V/nm, as shown in (c). In (d), the corresponding TEM micrograph of the apex of the Au nanocone is overlaid with the simulation from (b). For all simulations in this figure, the voltage applied to the Au nanocone was $V$ = 68 V, the distance between nanocone and cathode electrode was $d$ = 100 nm and the apex radius $r$ = 2.3 nm. The scale bar in (c) also applies to (a) and (b). In the main text, the values of the electric field from the FEM simulations were based on simulations as the one in (b).

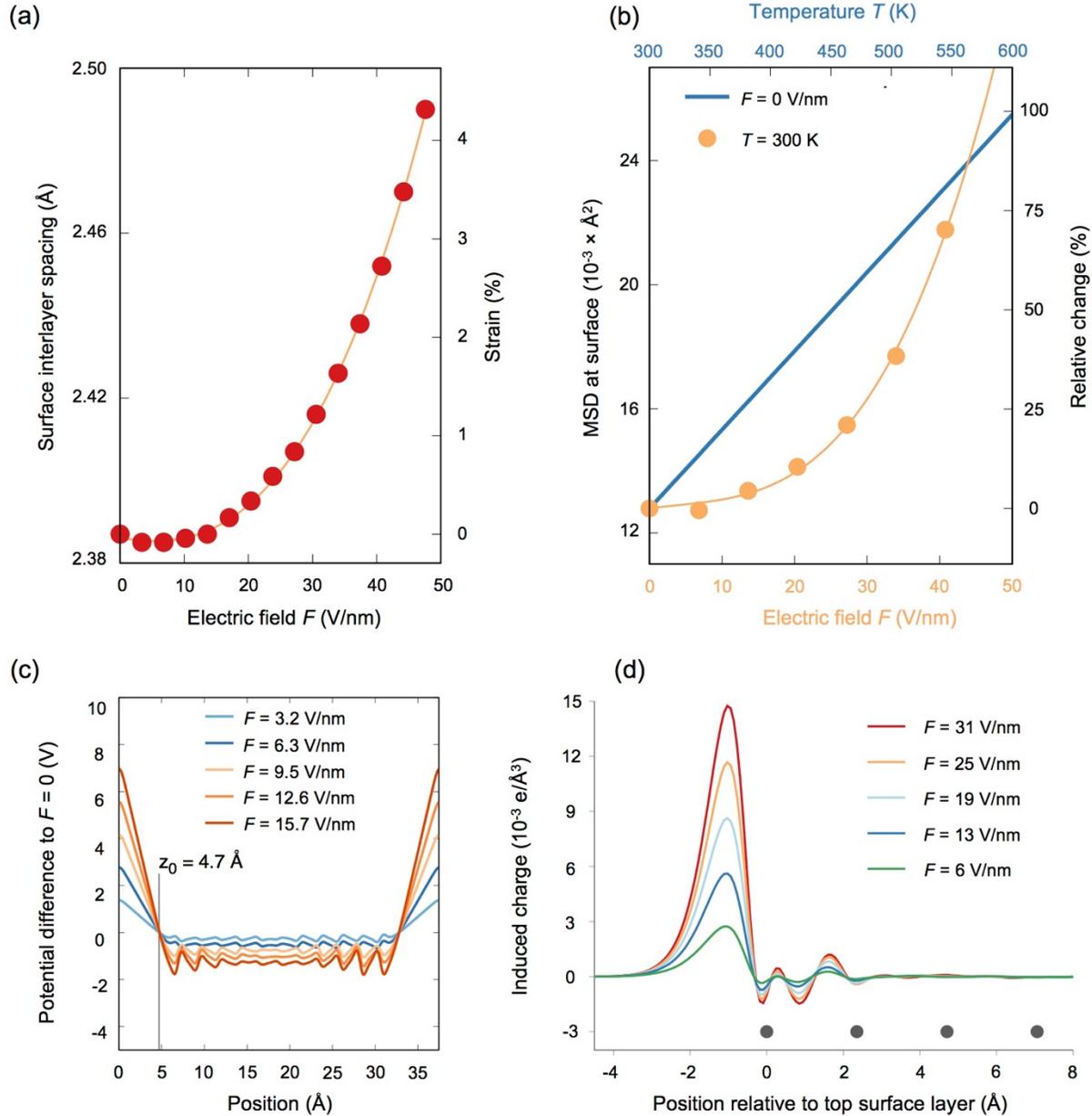

FIG. S4. *Ab initio* molecular dynamics simulations with density functional theory. (a) At low electric fields ($F$) the system responded by a small inward relaxation of the topmost layer, reflecting the ions to maximize the embedding in the electronic density. When the electric field reached 15 V/nm this was reversed as the ion-ion repulsion starts dominating and the topmost atomic layer relaxed outward instead. Eventually, at a field strength approaching 30 V/nm the spacing of the two top surface layers was about 1% larger than in the field-free case. (b) Mean square displacement (MSD) of the topmost surface layer as a function of temperature $T$ (blue curve) and electric field at $T = 300$ K (orange curve). (c) Induced charge at the surface of a (111) gold slab as a function of electric field. (d) The electric field induces a redistribution of charge at the surface, which served to screen the external field and was mostly localized to the first atomic layer. The data shown here were obtained for a {111} surface ignoring field induced atomic relaxation for clarity of presentation. The gray spheres at the bottom of the figure show the positions of the atomic layers.